\documentclass[aps,pre,reprint,showpacs,superscriptaddress]{revtex4-1}
\usepackage{graphicx}
\usepackage{dcolumn}
\usepackage{bm}
\usepackage{color}
\usepackage{lineno}
\begin{document}


\title{Monte-Carlo approach to calculate ionization dynamics of warm dense matter within particle-in-cell simulations}

\author{D. Wu}
\affiliation{State Key Laboratory of High Field Laser Physics, 
Shanghai Institute of Optics and Fine Mechanics, 201800 Shanghai, China}
\affiliation{Helmholtz Institut Jena, D-07743 Jena, Germany}
\author{X. T. He}
\affiliation{Key Laboratory of HEDP of the Ministry of Education, Center for Applied Physics and Technology, 
Peking University, 100871 Beijing, China}
\author{W. Yu}
\affiliation{State Key Laboratory of High Field Laser Physics, 
Shanghai Institute of Optics and Fine Mechanics, 201800 Shanghai, China}
\author{S. Fritzsche}
\affiliation{Helmholtz Institut Jena, D-07743 Jena, Germany}
\affiliation{Theoretisch-Physikalisches Institut, Friedrich-Schiller-University Jena, D-07743 Jena, Germany}

\date{\today}

\begin{abstract}           
A physical model based on a Monte-Carlo approach is proposed to calculate the ionization dynamics of warm dense matters (WDM) within particle-in-cell simulations, and where the impact (collision) ionization (CI), electron-ion recombination (RE) and ionization potential depression (IPD) by surrounding plasmas are taken into consideration self-consistently. When compared with other models, which are applied in the literature for plasmas near thermal equilibrium, the temporal relaxation of ionization dynamics can also be simulated by the proposed model. Besides, this model is general and can be applied for both single elements and alloys with quite different compositions. The proposed model is implemented into a particle-in-cell (PIC) code, with (final) ionization equilibriums sustained by competitions between CI and its inverse process (i.e., RE). Comparisons between the full model and model without IPD or RE are performed. Our results indicate that for bulk aluminium in the WDM regime, i) the averaged ionization degree increases by including IPD; while ii) the averaged ionization degree is significantly over estimated when the RE is neglected. A direct comparison from the PIC code is made with the existing models for the dependence of averaged ionization degree on thermal equilibrium temperatures, and shows good agreements with that generated from Saha-Boltzmann model or/and FLYCHK code. 
\end{abstract}

\pacs{52.38.Kd, 41.75.Jv, 52.35.Mw, 52.59.-f}

\maketitle

\section{Introduction}
Warm dense matter (WDM) \cite{wdm1, wdm2, wdm3}, with density $0.1$ to $10$ times that of solid and temperature $1$ to $100\ \text{eV}$, is commonly found in astrophysics as well as in high-energy density physics experiments \cite{icf2}. Until the present, however, the properties of WDM are not well understood and are difficult to predict theoretically. This is because neither the models of condensed-matter nor from high-temperature plasmas are well suited for describing the intermediate regime of WDM. 

Detailed information about the thermodynamic states, such as ionization distributions, is of importance in uncovering the involved physical mechanisms in WMD regime. Two widely applied models that predict an average ionization degree of atoms are Thomas-Fermi model \cite{TF} and Saha \cite{Saha} ionization model. Both of the models, however, assume that plasma conditions are near thermal equilibrium. For laser produced plasmas and intense beam solid interactions, where many of the involved physical processes take place at the sub-pico-second or pico-second scales \cite{lpi1, lpi2, lpi3, lpi4}, the equilibrium assumption is no longer correct. 
To account for the temporal evolution of the plasma ionization, an impact (collision) ionization (CI) model based on electron-ion collisional cross sections has been explored \cite{ci_pic_1, ci_pic_2, ci_pic_3}, which allows to calculate ionization values in a much more natural manner than equilibrium models. This model directly describes the inter-particle interactions in the plasmas and thus, accounts for the multi-particle nature of real plasmas. 
Although the CI model allows improvements in dealing with non-equilibrium plasmas, it is still not complete since it does not account for the inverse process, i.e., electron-ion recombinations (RE) \cite{ir_1, ir_2, ir_3, ir_4}. 
Besides, the ionization potential depression (IPD) should be taken into account when dealing with dense plasmas \cite{ipd_1, ipd_2, ipd_3, ipd_4, hedp.1, hedp.2}, however it is also ignored in the considered models \cite{ci_pic_1, ci_pic_2, ci_pic_3}. 

The main challenge to understanding the ionization of WMD is to incorporate self-consistently the non-linear behaviour in such strongly coupled dynamical systems, i.e., the matter's response to the surrounding plasmas and plasmas' response to the matter through CI, RE and IPD processes. To describe the ionization dynamics of WDM more systematically, we here propose and analyse a Monte-Carlo approach that can be configured and embedded into existing particle-in-cell (PIC) simulation codes. 
In this approach, we use a collection of macro-particles to describe a plasma or matter of finite ion density. 
Here, a macro-particle can be regarded as the ensemble of real particles, i.e., a group of particles with ``same'' mass, charge state, position and momentum. The electrons are classified moreover into bound and free ones, where the former are regarded as part of ions or atoms, and the latter are isolated as the surrounding plasmas. 
Since we consider a collection of a large number of particles and a pico-second temporal evolution of the system, the fine structures, such as sub-shell configurations, excitations and their inverse processes, are ignored in the present model. Only the dominant physical processes are taken into account, such as CI and RE. Furthermore, the IDP by the surrounding plasmas should also be taken into consideration. This is because it will lower the bounding energy of ions or atoms, which will then, in turn, affect both CI and RE processes. 

The paper is organized as follows. The physical model concerning CI, RE and IPD are introduced in Sec. II. In Sec. III, the model is embedded into a PIC simulation code. Comparisons between the full model and model without IPD or RE are performed and analysed. Dependence of averaged ionization degree on thermal equilibrium temperatures is obtained by the PIC code. Comparisons with results generated from Saha-Boltzmann model or/and FLYCHK code are made.
Summary and discussion are given in Sec. IV.     
    
\section{Physical model}
When temperature of plasma is high with the kinetic energy of free electrons exceeding the ionization potential of ions or atoms, there exists a possibility that the ion or atom will lose a bound electron by the colliding with energetic free electrons. Simultaneously, free electrons and charged ions also have the tendency to recombine together. Different from isolated atom or ion, the screening of plasmas would dramatically influence the atomic structure of ions or atoms that embedded in, resulting in the lowering of their bounding energies. The above three processes, CI, RE and IPD, are usually ignored in high temperature and low density plasmas. While in the WDM regime, these processes should be self-consistently taken into account.
In this section, a CI model based on electron-ion collisional cross sections, a RE model based on three-body-recombination and an IPD  model based on the pioneering works of Stewart and Pyatt are explored and implemented into an existing PIC simulation code. 
         
\paragraph{Impact ionization} Generally, a cross section of ionization can be derived by establishing an electron-ion (or atom) collisional pair and taking into account the energy of the incoming electron as well as the ionization state of the ion. 
The pioneering work was done by Lotz \cite{Lotz}, with the formula of the total cross section as follows
\begin{equation}
\label{total_cross}
\sigma^{\text{ci}}=\sum_{i=1}^{N}a_i q_i \frac{\ln{(E/P_i)}}{EP_i}[1-b_i\exp(-c_i(E/P_i-1))],
\end{equation} 
where $E$ is the energy of impact electron, $P_i$ is the binding energy of electron in the $i$-th sub-shell, $q_i$ is the number of equivalent electron in the $i$-th subshell, and $a_i$, of unit $10^{-14} \text{cm}^2 \text{eV}^2$, $b_i$ and $c_i$ are individual constants which are determined by experiment measurements or theoretical predictions. Ref.\ \cite{Lotz} also tabulates these constants of ionization cross sections, and this table is applied in our computations below. Furthermore, following Eq.\ (\ref{total_cross}), the ionization cross section among neighbouring levels, such as, Al-II to Al-III, can be formulated as follows,
\begin{equation}
\label{single_cross}
\sigma_i^{\text{ci}}=a_i q_i\frac{\ln{(E/P_i)}}{EP_i}[1-b_i\exp(-c_i(E/P_i-1))],
\end{equation}  
with $E\geq P_i$, where $P_i$ is the ionization potential from $i$ to $i$+$1$ charge state, such as Al$^{1+}$ to Al$^{2+}$. Let us note, however, the fine structure levels are ignored in Lotz's model, for which the ionization stage is treated as from the ground state to the next ground state. This assumption here is reasonable, as the fine structure levels are averaged out by the collection of large number of particles.  
Furthermore, the electron impact ionization cross section can also be calculated using the relativistic 
multi-configuration Flexible Atomic Code (FAC) \cite{fac}.
The impact ionization rate of ion or atom is 
\begin{equation}
\nu_i^{\text{ci}} = \int_{P_i}^{\infty}v_e\sigma_i^{\text{ci}}(E)f_e(E)dE,
\end{equation}
where $E$, $v_e$, and $f_e$ are energy, velocity and density of surrounding electrons with energy between $E$ and $E+dE$. In PIC simulations, the integral perform a summation over all electrons that reside within the same cell as the given ion of interest. 
The expression for $\nu_i$ in this form can be time-consuming as it requires a double loops over all ions and electrons in the cell. The idea presented in Ref.\ \cite{ci_pic_2} takes advantage of the specific scaling of the ionization cross section and electron velocity with energy, i.e., $\ln(E)/E$ and $\sqrt{E}$, respectively, whose product is not sensitive to $E$ and can be taken outside the integration. When replaced by their averaged values, the impact ionization rate takes the form 
\begin{equation}
\label{ci_rate}
\nu_{i}^{\text{ci}} =  \sigma_i^{\text{ci}}(\bar{E})\bar{v}n_e\ (\text{s}^{-1}),
\end{equation}
where $\bar{E}$, $\bar{v}$ and $n_e$ are the averaged energy, velocity and density of electrons in a cell.  
However, we have found that the above method tend to underestimate the ionization degree. When $\bar{E}<P_i$, ionization can not take place at all, as those energetic electrons, which play an important role in impact ionization, are averaged out in the above method. To improve the above method and simultaneously overtake the time-consuming double loops, our idea is as follows: i) a loop over electrons generates the average electron energy $\bar{E}$; ii) preparing three arrays, $\bar{E}_m$, $\bar{n}_{em}$ and $\bar{v}_m$ containing the averaged energy, density and velocity of electrons with their energies spanned by $\bar{E}_m$ and $\bar{E}_m+dE$ (the array step and maximal energy are assumed to be $0.25\times\bar{E}$ and $5\times\bar{E}$);
iii) a loop over electrons is performed again to fulfil the arrays; iv) ionization rate for each ion in a cell is calculated by the following formula,      
\begin{equation}
\label{ci_rate_new}
\nu_{i}^{\text{ci}} =  \sum_{m=0}^{20}\sigma_i^{\text{ci}}({\bar{E}_m})\bar{v}_m\bar{n}_{em}\ (\text{s}^{-1}).
\end{equation}
The ionization probability of the ion of interest is $p_{i}^{\text{ci}}=1-\exp(-\nu_{i}^{\text{ci}}\delta t)$, where $\delta t$ is the  time step of PIC simulation. We increase the ionization degree by one unit for each ion and simultaneously put in an electron with the same position, velocity and weight as its host ion, when condition $r>p_i^{\text{ci}}$ is satisfied, where $r$ is the computer generated random number. To ensure that the energy remains conserved in the computations, we reduce local kinetic energy by distributing a momentum reduction to all local electrons, which is equivalent to the ionization energy.

\paragraph{Electron-ion recombination} Usually, the ionization balance of a plasma is determined by the competing processes of CI and RE, as well as various excitation/de-excitation processes. In particular, the recombination of electrons and ions takes place mainly by three different reaction modes, the dielectronic (D-RE), radiative (R-RE) and three-body recombinations (TB-RE), respectively \cite{ir_1}. As we have analysed, in our model only ground state of ions and atoms are concerned, the contributions of D-RE are averaged out. Note that, R-RE is the inverse process of direct photo-ionization, while TB-RE is the inverse process of electron impact ionization. R-RE process is known to predominantly fill the low-lying Redberg states, while TB-RE is mainly responsible in rapidly bringing the high Rydberg states into equilibrium \cite{ir_1}. Thus, the contributions of recombinations in our cases mainly arise from the TB-RE process, with $e+e^{'}+\text{A}^{Z} \rightarrow \text{A}^{ZM}+e^{''}$, where $ZM=Z-1$ with $Z$ of the ion charge state. In the TB-RE, the excess energy released by the recombining electron is carried away by the outgoing electron $e^{''}$, so that the TB-RE does not involve any emission of photons. 

Expression of TB-RE rate formula has a strong dependence on the relying impact ionization formula. 
Let us consider the detailed balance equation of species with ionization charge state $i$ and $i$+$1$,
\begin{equation}
\label{ir_rate}
\frac{\partial n_i}{\partial t} = \nu_{i\text{+}1}^{\text{re}} n_{i+1} - \nu_i^{\text{ci}}  n_i,
\end{equation} 
where $n_i$ is the density of ions, $\nu_{i+1}^{\text{re}}$ is the three-body recombination rate and $\nu_{i}^{\text{ci}}$ is impact ionization rate. In order to relate these rate coefficients, one observes that at the recombination-ionization equilibrium, 
we have $\nu_{i\text{+}1}^{\text{re}} n_{i\text{+}1}= \nu_i^{\text{ci}} n_i$. 
As the ionization equilibrium can be well described by the Saha-Boltzmann Equation \cite{ir_1}, 
\begin{equation}
\frac{n_e n_{i\text{+}1}}{n_i}=\frac{g_e g_{i\text{+}1}}{g_i}(\frac{2\pi m_e k_BT_e}{h^2})^{3/2} \times \exp{(-\frac{P_{i}}{k_BT_e})}
\end{equation}
one can obtain, 
\begin{equation}
\label{re_rate}
\nu_{i\text{+}1}^{\text{re}}=\frac{g_i}{g_e g_{i\text{+}1}}(\lambda_e)^3n_e\times\exp{(\frac{P_{i}}{k_BT_e})}\times\nu_i^{\text{ci}},
\end{equation}
where $\lambda_e=\sqrt{h^2/2\pi m_e k T_e}$ is the thermal electronic de Broglie wavelength,
$g_e$ and $g_i$ are the statistical weights, and $\nu_i^{\text{ci}}$ is the ionization rate as shown in Eq. (\ref{ci_rate_new}).  

According to Eq.\ (\ref{re_rate}), the recombination rate is increased dramatically in low temperature and high density plasma environment. Note that all the TB-RE formulas \cite{ir_2, ir_3, ir_4} exhibit this behaviour, except for the slightly different numerical factors. In PIC simulations, the electron temperature $T_e$, and electron density $n_e$ can be generated by a loop over the electrons in each computational cell at every time step. 
Then Eq.\ (\ref{re_rate}) is applied for each ion resides in the same cell. 
The recombination probability is $p_{i}^{\text{re}}=1-\exp(-\nu_{i}^{\text{re}}\delta t)$, where $\delta t$ is the simulation time step. We decrease ionization degree by one unit for each ion, whenever the random number $r$ satisfies $r>p_i^{\text{re}}$. Again to ensure that the energy remains conserved, the local kinetic energy, equivalent to the ionization energy, is increased, through a similar way as we have done in impact ionizations, by distributing a momentum modification to all local electrons. 
To ensure the conservation of remain particles, the local plasma density, equivalent to recombinations, is reduced by distributing a weight modification to all local free electrons.  

\begin{table*}
\caption{\label{table1} Ionization potential of aluminium atom and ions from NIST \cite{nist} as implemented in our model.}
\begin{ruledtabular}
\begin{tabular}{ l  l  l  l  l  l  l l  l  l  l  l  l  l}
Al & 1 & 2 & 3 & 4 & 5 & 6 & 7 & 8 & 9 & 10 & 11 & 12 & 13 \\ 
eV & 5.980 & 18.80 & 28.40 & 119.9 & 153.8 & 190.4 & 241.4 & 284.5 & 330.1 & 398.5 & 441.9 & 2085. & 2300. \\
\end{tabular}
\end{ruledtabular}
\end{table*}

\paragraph{IPD by surrounding plasmas} The calculation of both impact ionization and electron-ion recombination requires values of ionization potentials, which, in principle, can be generated or obtained from data bases of National Institute of Standard and Technology (NIST). The ionization potential of aluminium atom (Al I) and ions are listed in Table \ref{table1}, which are calculated based on the isolated atom or ion model. However in a plasma of finite density and temperature, the ionization potential of a given ion is influenced not only by its own bound electrons but also by the surrounding free electrons, which, in turn, will affect both impact ionization and recombination processes. 
Therefore, the phenomenon of ionization-potential depression for ions embedded in the plasma are of crucial importance for modelling atomic processes within dense plasmas. {We here refer to the theory of IPD as introduced by Stewart and Pyatt \cite{ipd_2}, which is widely used in literatures of plasma and atomic physics calculations, including FLYCHK \cite{flychk1, flychk2} code.} The model yields ion-sphere and Debye-Huckel potential models as approximate limiting cases and could provide results over essentially the entire range of temperature and densities of plasmas. Let us here consider an ion (or atom), $i$, fixed in a sea of free electrons and ions at kinetic temperature $T_e$. The free electrons are described by relativistic Fermi-Dirac statistics and the ions by non-relativistic Maxwell-Boltzmann statistics. For such a distribution of plasma electrons, the average electro-static potential near $i$ can be evaluated by Poisson equations. It is this potential that cause the IDP of the ion. The contributions of bound electrons to IPD are excluded, since they are already present in the isolated ion. 

Following the work of Stewart and Pyatt, 
the lowering of ionization potential is described by,
\begin{equation}
\label{IDP_formula}
\Delta P=\{[3(Z+1)K+1]^{2/3}-1\}T_e/2(Z+1),
\end{equation} 
where $T_e$ is temperature of free electrons (plasmas), and $K=Z e^2/\lambda_d T_e$ with $\lambda_d=\sqrt{T_e/4\pi Z n_e}$ represents the Debye length of free electrons. For small $K$ values, according to Eq.\ (\ref{IDP_formula}), $\Delta P$ is reduced to $Z e^2/\lambda_d$ which is the limit of Debye-Huckel model. When $K$ is large, $\Delta P$ equals to $3Z e^2/2a$, which is the limit of ion-sphere model, with $a=\sqrt[3]{3Z/4\pi n_e}$ representing the radius of ion-sphere. 
For high density plasmas, the IDP would have a significant effect on lowing of ionization potential. For example, $\Delta P$ of Al VII (with the isolated ionization potential $240\ \text{eV}$) can be as large as $100\ \text{eV}$ for bulk aluminium ($2.7\ \text{g}/\text{cm}^3$) with temperature $T_e$ below $300\ \text{eV}$. 
{Note the IPD calculation by itself is still open in the WDM research.  
For going beyond such a semi-empirical treatment, a rigorous way of dealing with IDP is through multi-body-quantum-mechanical methods \cite{ipd_3, ipd_4}. We have compared the values generated from Stewart and Pyatt's formula with that from references \cite{ipd_3, ipd_4}. Results indicate that both calculation methods exhibit similar behaviour, though with slightly different numerical values.} 
In PIC simulations, electron temperature $T_e$ and density $n_e$ can be generated by a loop over electrons in each computational cell, attached to which the Debye length $\lambda_d$ is evaluated. Using Eq.\ (\ref{IDP_formula}) and isolated ionization potential value from NIST data bases, the modified ionization potential, i.e., $P-\Delta P$, is updated for each ion at every computational cell per time step.  

\section{Applications}
The above three processes are embedded in a recently extended version of PIC code based on LAPINE \cite{code1}. This is a parallel high-order-scheme PIC code written in C++ language, capable of performing 1-D, 2-D and 3-D simulations, with which the tunnelling ionization \cite{code2}, relativistic binary collisions \cite{code3}, radiation reaction and photon emission in quantum electrodynamics regime \cite{code4} have already been implemented in by one of us. In this section, we will present several case studies of the ionization dynamics of bulk aluminium (single) and aluminium carbide (alloy). {Let us note that the initially assumed charge state does not depend on the initial temperature in the following calculations, and that the free electron temperature is taken from a reasonable guess. The dependence of averaged ionization degrees on temperatures can only be established at (final) thermal equilibrium, after a reasonable relaxation time.}  

\begin{figure}
\includegraphics[width=8.50cm]{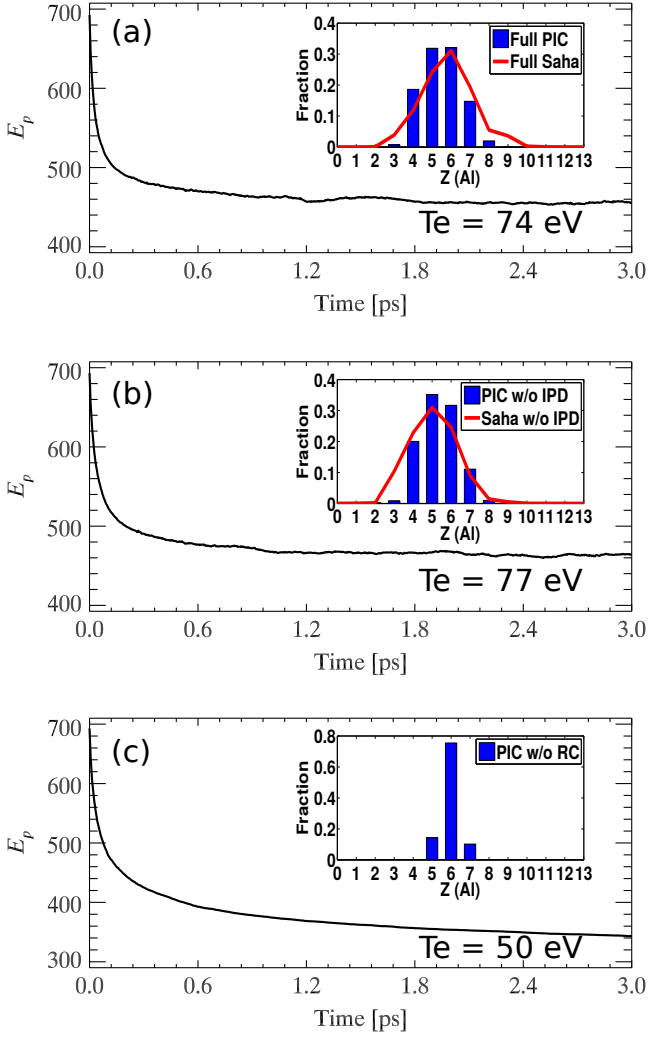}
\caption{\label{fig1} (color online) (a) The total plasma energy (A. U.), with the full model by summarizing over all free electrons within a computational cell, as a function of time, with initial plasma temperature $150\ \text{eV}$ and pre-defined charge state $4+$. (b) The same as shown in (a), but with the model excluding IPD. (c) The same as shown in (a), but with the model excluding RE. The inlets over (a) (b) and (c) are the corresponding final ionization distributions of aluminium after $3\ \text{ps}$ relaxation. The red line covered on the inlets are the ionization distributions of aluminium calculated by Saha-Boltzmann Equation with defined temperature, (a) $T_e=74\ \text{eV}$ and (b) $T_e=77\ \text{eV}$ also excluding IPD.}
\end{figure}

\begin{figure}
\includegraphics[width=8.50cm]{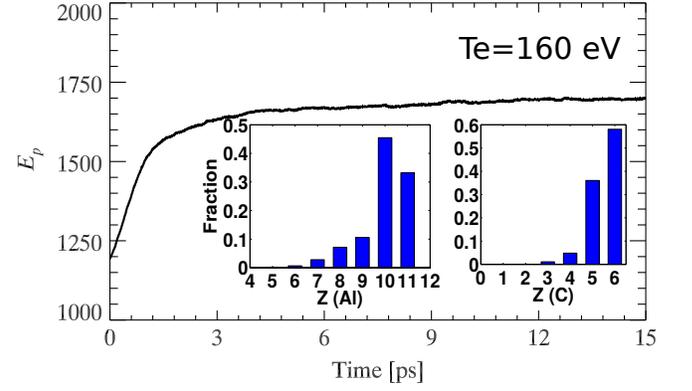}
\caption{\label{fig2} (color online) The total plasma energy (A. U.), with the full model by summarizing over all free electrons within a computational cell, as a function of time, with the initial temperature of aluminium carbide $100\ \text{eV}$ and pre-defined charge states of $11+$ for aluminium and $6+$ for carbon. The inlet is the final ionization distributions of aluminium and carbon after $15\ \text{ps}$ relaxation.}
\end{figure}

The density of bulk aluminium in our case studies is $2.7\ \text{g}/\text{cm}^3$, thus, the aluminium ion density is $6.6\times10^{22}/\text{cm}^3$. The initial aluminium charge state is assumed to be $4+$, and the initial free electron temperature is set to $150\ \text{eV}$. As a benchmark of the ionization dynamics, we consider only a few computational cells, connected by periodic boundaries conditions, with each cell contains $200$ ion macro-particles and $200$ electron macro-particles initially. The grid size of PIC simulation is $0.01\ \mu\text{m}$ and time step is set to $0.02$ fs. {In the simulations, we have also taken into account the collisions between electrons, ions, and electron-ion.} To figure out the influence of IPD and RE, three sets of simulations are run simultaneously. PIC simulations with full model (CI+IPD+RE), model without IPD and model without RE are present in Fig.\ \ref{fig1} (a) (b) and (c). 
Fig.\ \ref{fig1} (a) shows the total plasma energy (A. U.), with the full model by summarizing over all free electrons within a computational cell, as a function of time. Fig.\ \ref{fig1} (b) and (c) are the same as shown in (a), but with the model excluding IPD and RE, respectively. 
Following the energy history, at initial time, the CI rate of aluminium is larger than RE. The former one would reduce the plasma energy and increase the averaged ionization degree as a function of time.
{Compared with Fig.\ \ref{fig1} (a), we found that after $6$ ps relaxation, the averaged ionization degree is lowered when excluding the IPD, which is $\bar{Z}=5.82$ with $T_e=74\ \text{eV}$ (a) v.s. $\bar{Z}=5.05$ with $T_e=77\ \text{eV}$ (b). From the comparison with Fig.\ \ref{fig1} (a) and (c), we found that after $6$ ps relaxation, the averaged ionization degree is significantly over estimated when excluding the RE process. Note that Fig.\ \ref{fig1} (c) also, in principle, represent the results of existing PIC code \cite{ci_pic_1, ci_pic_2, ci_pic_3}, with which only CI is taken into account.} As presented in Eq.\ (\ref{re_rate}), RE would become a dominant process for ions embedded in plasmas of high density and moderate temperatures. 

Our model is general and can be applied for both single elements and alloys with quite different compositions. 
The aluminium carbide, chemical formula Al$_4$C$_3$, is a carbide of aluminium with density $2.36\ \text{g}/\text{cm}^3$. 
The simulation set is the same as shown in Fig.\ \ref{fig1}, but with an additional species carbon. 
Fig.\ \ref{fig2} shows the total plasma energy (A. U.), with the full model by summarizing over all free electrons within a computational cell, as a function of time, with the initial temperature of aluminium carbide $100\ \text{eV}$ and pre-defined charge states of $11+$ for aluminium and $6+$ for carbon. 
As shown in Fig.\ \ref{fig2}, thermal equilibrium is reached after $15$ ps relaxation. The inlet is the final ionization distributions of aluminium and carbon with thermal equilibrium temperature $160\ \text{eV}$.

\begin{figure}
\includegraphics[width=8.50cm]{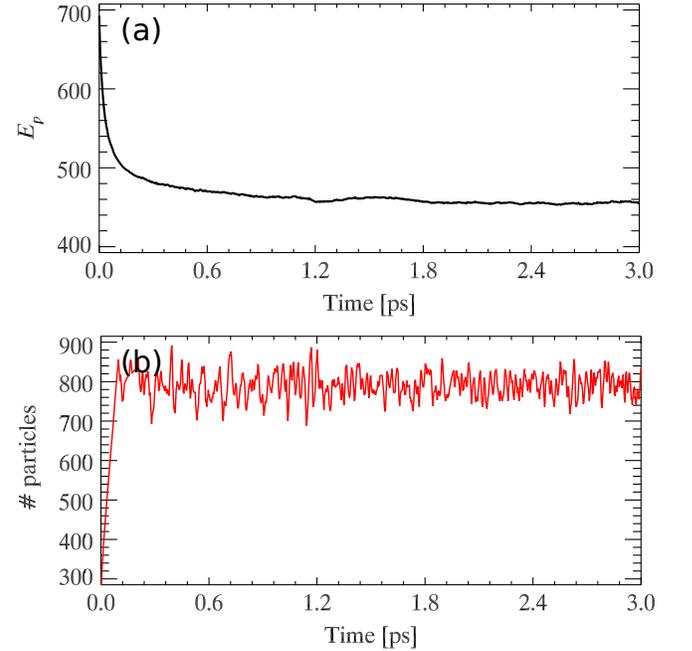} 
\caption{\label{fig3} (color online) (a) The total plasma energy (A. U.), with the full model and merging particle technique by summarizing over all free electrons within a computational cell, as a function of time, with initial plasma temperature $150\ \text{eV}$ and pre-defined charge state $4+$. (b) The corresponding temporal fluctuation of the number of macro-particles.}
\end{figure}

\begin{figure}
\includegraphics[width=8.50cm]{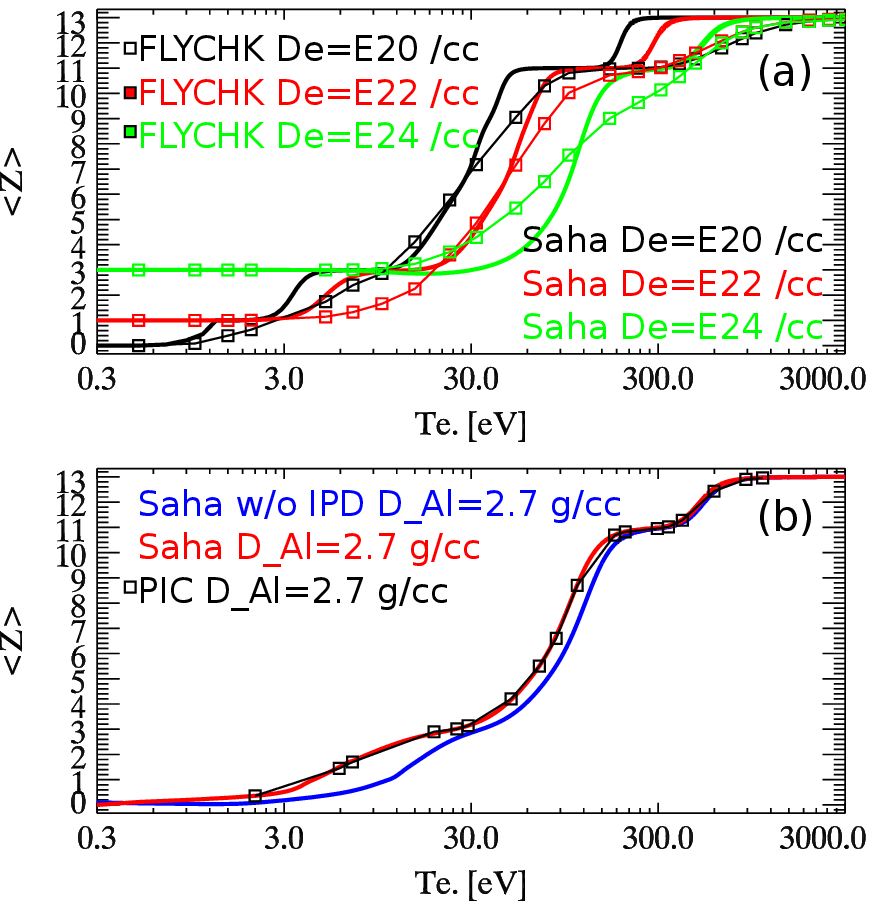}
\caption{\label{fig4} (color online) The averaged ionization degree of bulk aluminium as a function of plasma temperature. 
(a) Blue, red and green lines (square) are the results calculated by Saha-Boltzmann Equation (FLYCHK code), with fixed electron density of $10^{20}\ \text{cm}^{-3}$, $10^{22}\ \text{cm}^{-3}$ and $10^{24}\ \text{cm}^{-3}$. 
(b) Red and green lines are the results calculated by Saha-Boltzmann Equation with updated numerical scheme, including IPD and excluding IPD, with fixed aluminium density $2.7\ \text{g}/\text{cm}^3$. Black square is picked up from the equilibrium states calculated by our PIC code with full model.}
\end{figure}

In our model, different strategies were used to ensure that the total number of particles remain conserved. The reduction of electrons due to recombination is through distributing a weight modification to all local free electrons, which does not change the number of macro-particles. While the increase of electron due to impact ionization is through placing \textit{new} macro-particles to the cell of interest. The major computational effort in the simulation arises from the number of macro-particles. To solve this problem, a particle-merging technique is configured and applied. Considering two electrons with position $\bm{r}_a$ and $\bm{r}_b$, momentum $\bm{p}_a$ and $\bm{p}_b$, gamma factor $\gamma_a$ and $\gamma_b$, as well as weight $w_a$ and $w_b$, we have the merging weight as $w=w_a+w_b$, merging position as $\bm{r}=(w_a\bm{r}_a+w_b\bm{r}_b)/w$, merging gamma factor as $\bm{\gamma}=(w_a\gamma_a+w_b\gamma_b)/w$, and the merging momentum as $\bm{p}=(w_a\bm{p}_a+w_b\bm{p}_b)/w$. In practice, the equation, $\gamma=\sqrt{p^2+1}$, is not always satisfied for the \textit{merged} particles. To solve this problem, a coefficient of $\eta=\sqrt{(\gamma^2-1)/p^2}$ is multiplied to replace the old merging momentum, with $\bm{p}=\eta\bm{p}$. 
The case shown in Fig.\ \ref{fig1} (a) is re-run by including the merging-particle technique. 
Fig.\ \ref{fig3} (a) shows how the total plasma energy evolves in time, while Fig.\ \ref{fig3} (b) displays the corresponding number of macro particles. Both the energy evolution and final equilibrium shown in Fig.\ \ref{fig3} (a) is exactly the same 
as shown in Fig.\ \ref{fig1} (a). In simulations, merging can be set to take place at pre-defined times when satisfying pre-defined conditions. In the case simulation shown in Fig. \ref{fig3}, merging is set to take place at every $100$ time steps when number of macro-particles in a cell exceeding $1000$ ($200$ macro-particles are placed in a cell initially). To make this technique numerical stable, we would suggest the threshold of merging to be set to $3\sim5$ times the initial number of macro-particles in a cell.   
As we can see, the dropping of the total number of macro particles does not affect the energy evolution or final equilibrium. By using this technique, the simulation burden can be dramatically released.    

{
At present, we have compared with model calculation \textit{with} and \textit{without} the IPD and RE, a comparison that refers to the PIC code itself. In this section, a direct comparison with equilibrium models is made. As we have mentioned, the ionization equilibrium is described by the Saha-Boltzmann Equation, with
${n_e n_{i\text{+}1}}/{n_i}=({g_e g_{i\text{+}1}}/{g_i})({2\pi m_e k_BT_e}/{h^2})^{3/2}\times\exp{(-{P_{i}}/{k_BT_e})}$,
where $n_e$ ($n_i$), $g_e$ ($g_i$), $P_i$ and $T_e$ are electron (ion) density, statistical weights, ionization potential and thermal equilibrium temperatures. Note that $P_i$ can be obtained from the NIST database \cite{nist}. While in WDM regime, as we have analysed, $P_i$ should be corrected by taking into account IPD, which can be calculated by Stewart and Pyatt's formula. To solve the above Saha-Boltzmann Equation, a natural way is to i) normalize the above equation by $n_e$, $\widetilde{n_i}=n_i/n_{\text{e}}$, ii) establish an iterative scheme, iii) guess a initial values of $\widetilde{n_1}$, $\widetilde{n_2}$, $\widetilde{n_3}$... and iv) loop the iterative scheme until the required resolution is satisfied. Results of solving Saha-Boltzmann Equation by this method are shown in Fig.\ \ref{fig4} (a). The solid lines show the averaged ionization of aluminium as functions of electron density and temperatures, whereas the black, red and green lines represent the ones with electron densities fixed at $10^{20}\ \text{cm}^{-3}$, $10^{22}\ \text{cm}^{-3}$ and $10^{24}\ \text{cm}^{-3}$, respectively. In Fig.\ \ref{fig4} (a), we also present results obtained from FLYCHK, 
with which the ionization calculation is also based on the Saha-Boltzmann Equation. Both methods indicate that for fixed electron density at $10^{20}\ \text{cm}^{-3}$, averaged ionization degree is close to zero at low temperatures (room temperature) limit, while it becomes $1+$ or $3+$ when electron density is fixed at $10^{22}\ \text{cm}^{-3}$ or $10^{24}\ \text{cm}^{-3}$. Actually this non-zero averaged ionization degree is due to IPD. At high density and low temperature limit, the value of IPD can be even larger than the isolated ionization potential, which will free the $3$p$^1$ (and $3$s$^2$) electron. 

For aluminium of density $2.7\ \text{g}/\text{cm}^3$, i.e., $n_{\text{Al}}=6.7\times10^{22}\ \text{cm}^{-3}$, 
at low temperature limit, averaged charge degree $3+$ corresponds to a plasma of density $2.0\times10^{23}\ \text{cm}^{-3}$, 
which is consistent with the green (square) line in Fig.\ \ref{fig4} (a). However we still notice that averaged charge degree $0.01+$ corresponds to a plasma of density $10^{20}\ \text{cm}^{-3}$, which is, in contrast, consistent with the black (square) line in Fig.\ \ref{fig4} (a). Thus it is hard for us to judge the averaged ionization degree of a bulk aluminium at low temperature limit. 
The ``double-value'' comes from the numerical scheme in solving the Saha-Boltzmann Equation. 
In the first step of the numerical scheme, we normalize $n_i$ by $n_e$. Although it is a quite natural way of doing so, in real situations, $n_{\text{Al}}$ is fixed instead of $n_e$. 

Here we update the numerical scheme, with i) normalizing $n_i$ by $n_{\text{Al}}$, $\widetilde{n_i}=n_i/n_{\text{Al}}$ and ii) adding a new constraint condition $\sum_{i=0}^{i=13}{\widetilde{n_i}}=1$. For aluminium of fixed density $n_{\text{Al}}=6.7\times10^{22}\ \text{cm}^{-3}$, the averaged ionization degree as a function of temperature is present in Fig.\ \ref{fig4} (b). Red and blue lines correspond to the cases including and excluding IPD. Results indicate that, at low temperature, i) the averaged ionization degree of bulk aluminium is indeed close to zero, and ii) the averaged ionization degree when including IPD effect is indeed higher than excluding this effect;
}
In Fig.\ \ref{fig1} (a) and (b), the ionization distributions calculated by Saha-Boltzmann Equation with updated numerical scheme is present in the red curves covered on the inlets, showing good consistence with the PIC calculations. Furthermore, following the same routine as introduced by Fig.\ \ref{fig1}, the dependence of averaged ionization degree on thermal equilibrium temperatures covering a large variation is obtained by the PIC code, as shown in black squares in Fig.\ \ref{fig4} (b), also showing good consistence with results from Saha-Boltzmann Equation.     

\section{Conclusions and discussions}
In summary, a physical model based on Monte-Carlo approach is proposed to calculate the ionization dynamics of WDM within PIC simulations, where CI, RE and IPD by surrounding plasmas are taken into consideration self-consistently. When compared with other models, which are applied in the literature for plasmas near thermal equilibrium, the temporal relaxation of ionization dynamics can also be simulated by the proposed model. The proposed model is implemented into a PIC code, with (final) ionization equilibriums sustained by competitions between CI and RE. Comparisons between the full model and model without IPD or RE are performed. Results indicate that for bulk aluminium in the WDM regime, i) the averaged ionization degree when including IPD effect would be higher than excluding this effect; and ii) the averaged ionization degree is significantly over estimated when excluding RE effect. As a direct comparison with the existing models, dependence of averaged ionization degree on thermal equilibrium temperatures is obtained by the PIC code, showing good agreements with that generated from Saha-Boltzmann model or/and FLYCHK code. 

In our model, the explicit RE formula is determined by the relying impact ionization formula and Saha-Boltzmann Equation. The good agreements between values from PIC simulation at (final) thermal equilibrium and results from Saha-Boltzmann Equation are thus guaranteed by the proposed model.   

\begin{acknowledgments}
D. Wu wishes to acknowledge the financial support from German Academic Exchange Service (DAAD) and China Scholarship Council (CSC),
also thanks H. Xu at National University of Defence Technology (China), B. Goswami, J. W. Wang and S. Z. Wu at Helmholtz Institut-Jena (Germany) and S. X. Luan at Shanghai Institute of Optics and Fine Mechanics (China) for fruitful discussions. 
\end{acknowledgments}

{}

\end{document}